\documentstyle[aps,preprint]{revtex}  
\begin{document}                
\title{$\nu =1/2$ Bilayer Tunnelling}
\author{D. Lidsky}
\address{Department of Physics, Rutgers University, Piscataway, NJ 08855} 
\maketitle
\begin{abstract}

Tunneling of electrons between two $\nu = 1/2$ quantum Hall parallel planes is
studied.  In order to calculate the physical electron Green's function, the 
Chern-Simons gauge field formalism is used, both in a perturbative many-body 
calculation and in a semiclassical calculation. In contrast to recent 
experiments, no gap in the physical electron density of states is found.  A 
conjecture is made that either perturbation theory about the mean field Fermi 
liquid ground state is inadequate, or that disorder within the interplanar 
barrier plays a role.

\end{abstract}
Experimental measurements of tunnelling between two quantum Hall layers reveal
a gap-like current-voltage characteristic, $I=I_{0}\exp (-\Delta /V)$.    
For odd denominator filling factors, this behavior is qualitatively, but
not quantitatively, 
explainable in terms of single plane transport properties.  In particular,
although the single plane experiments for $\nu = 1/3$ show a gap only of order
  2K   \cite{boebinger} \cite{prange}, the bilayer
tunneling measurements show a much larger gap of order 100K \cite{eis}.  
For even denominator filling factors, there is not even qualitative similarity
between the single layer and the bilayer spectrum.  The single layer 
$\nu = 1/2$ quantum Hall state  possesses a Fermi surface.
Nevertheless, tunnelling experiments have shown gap-like behavior over a 
broad range of external magnetic fields, corresponding to both odd and 
even filling factors \cite{eis}.

	He, Platzmann, and Halperin have
calculated the tunnelling rate and have gotten good agreement with experiment
\cite{he}.  They calculated the one-electron Green's function by analogy to
the x-ray edge problem, in which a localized deep hole interacts with some
bosonic excitations.  In order to make the analogy, they assume that the
electron which tunnels into the adjacent plane is localized because it sees
the strong external magnetic field.  On the other hand all the other electrons
have been transformed into Chern Simons fermions which see no effective
magnetic field.    

	In the present work, we express all electrons, in particular the
tunnelling electron, in terms of Chern-Simons fermions attached to two
flux tubes.  
	
	Since the same behavior was observed experimentally over a wide
range of filling factors, the explanation must be universal.  In other
words, some may question the point of a theory that applies to
$\nu = 1/2$ only.  Indeed, several authors have explained the result 
classically as the problem of overcoming the Coulomb blockade, or as the 
problem of an extra electron attempting to merge into a rigid Wigner crystal
\cite{efros}.  However, these classical explanations leave certain questions
unanswered.  The Coulomb blockade explanation implies incompressibility.
But, experimentally, the single plane $\nu = 1/2$ state has finite 
conductivity and thus is compressible.  Likewise, the Wigner crystal is
not observed for $\nu = 1/2$.  Even if those contradictions can be resolved, 
it is still important to check
the validity of the Chern-Simons formalism of the half-filled Landau level.
We will see that perturbative corrections to the mean field Fermi liquid 
ground state do not lead to a decaying
electron Green's function and consequently do not reproduce even 
qualitatively the experimental results.

\section{Physical Electron Green's Function}

\newcommand{\be}{\begin{equation}}
\newcommand{\ee}{\end{equation}}
\def\ba{\begin{eqnarray}}
\def\ea{\end{eqnarray}}

  Following \cite{hlr},\cite{kz}, a physical electron is transformed into
a Chern-Simons fermion as follows
\begin{equation}
\psi_{e}^{\dag}(r) = \psi^{\dag}(r)
\exp[i\tilde{\phi}\int
d^{2}r\prime arg(r-r\prime)\rho(r\prime)]
\end{equation}
In the present work, it will prove convenient to rewrite the phase in terms 
of the the contour integral of the Chern-Simons gauge field,$\vec{a}$.
\begin{equation}
\psi_{e}^{\dag}(r) = \psi^{\dag}(r)
\exp[-ig\int_{C_{r}}
d\vec{r\prime}\cdot \vec{a}(\vec{r\prime})] \label{eq:Chern-Simons sphase}
\end{equation}
$g$ is inserted for convenience.  Since we will soon do perturbation
theory in powers of $a$, $g$ can be thought of as a small parameter which
will later be set equal to one.
The contour of integration,$C_{r}$, is determined by the convention that
arguments of vectors in the plane be measured with respect to the x-axis.
The contour goes partially around the periphery of the sample and then goes
to the point $r$ along the negative $x$-axis (Figure 1).
Since the physical problem is rotationally invariant, no physical quantity
should depend on the choice of $C_{r}$.  It can be shown that a different
convention for $C_{r}$ is equivalent to a gauge transformation and thus is
unmeasurable.

We write the  physical electron Green's function as a functional
integral over CS fermions and CS gauge fields
\begin{eqnarray}
\langle \psi_{e}(t)  \psi_{e}^{\dag}(0) \rangle & = &
\int D\psi D\psi^{\ast} Da_{i}Da_{0}e^{-S_{0}}e^{-gS_{int}}
 \psi(t)\psi^{\dag}(0) \nonumber \\
                                                &   & \exp[-i\int_{C_{0}}
d\vec{r\prime\prime}\cdot g\vec{a}(\vec{r\prime\prime},t)]
\exp[i\int_{C_{0}}
d\vec{r\prime}\cdot g\vec{a}(\vec{r\prime},0)] /Z  \label{eq:funcint}
\end{eqnarray}
The interaction between the fermion and the gauge field is
\begin{equation}
S_{int} = -ig\int d\tau dr^{2}\psi^{*}\vec{a} \cdot \nabla \psi
\label{eq:interaction}
\end{equation}

\section{Perturbation Theory}
 Notice
that (~\ref{eq:Chern-Simons sphase}) and (~\ref{eq:interaction}) each contain the
small parameter, $g$.
When doing a perturbative calculation, expanding in $g$, there are terms
from the Chern-Simons phase (~\ref{eq:Chern-Simons sphase})  and
 terms from the interaction (~\ref{eq:interaction}).  Some care must
be given when applying diagrammatic perturbation theory; in particular, not
all disconnected diagrams cancel.  

 The disconnected diagrams coming
from the Chern-Simons phase alone contribute an overall multiplicative
factor, given by the linked cluster expansion, to the on-site physical
electron Green's function.  First let us inspect the lowest order $g^{2}$
terms involving the contour phase only (~\ref{eq:Chern-Simons sphase}).
\begin{eqnarray}
&   &\int D\psi D\psi^{\ast} Da_{i}Da_{0}e^{-S_{0}}
 \psi(t)\psi^{\dag}(0) \nonumber \\
                                                &   &
 [1-ig\int_{C_{0}} dr\prime_{i} a_{i}(\vec{r\prime},t)
-\frac{1}{2}g^{2}\int_{C_{0}}dr\prime_{i}\int_{C_{0}}dr\prime\prime_{j}
 a_{i}(\vec{r\prime},t)a_{j}(\vec{r\prime\prime},t)] \nonumber \\
                                                &   & 
 [1+ig\int_{C_{0}} dr\prime\prime_{j} a_{j}(\vec{r\prime\prime},0)
-\frac{1}{2}g^{2}\int_{C_{0}}dr\prime_{i}\int_{C_{0}}dr\prime\prime_{j}
 a_{i}(\vec{r\prime},0)a_{j}(\vec{r\prime\prime},0)]
 /Z  \nonumber \\
  & = & \langle \psi(t)  \psi^{\dag}(0) \rangle
[-\int_{C_{0}}dr\prime_{i}\int_{C_{0}}dr\prime\prime_{j}
{\cal D}_{ij}(r\prime -r\prime\prime,0)
+ \int_{C_{0}}dr\prime_{i}\int_{C_{0}}dr\prime\prime_{j}
{\cal D}_{ij}(r\prime -r\prime\prime,t)]\nonumber \\
  & = & G^{0}(t)\int_{C_{0}}dr\prime_{i}\int_{C_{0}}dr\prime\prime_{j}
\int(d\omega dk){\cal D}(\omega,k)
e^{ik\cdot(r\prime - r\prime\prime)}[e^{-i\omega t}-1]
(\delta_{ij}-  \frac{k_{i}k_{j}}{k^{2}})
\nonumber \\
  & \approx & G^{0}(t)\int(d\omega dk){\cal D}(\omega,k)
[e^{-i\omega t}-1]/k^{2}
\nonumber \\
  & \sim & G^{0}(t)(1-1/\sqrt{\epsilon_{F}t})
\end{eqnarray}
Diagrammatically, the fourth line of the previous expression corresponds to 
Figure 2.
We can sum up all such disconnected diagrams using the linked cluster expansion
\begin{equation}
G(t)  \sim  G^{0}(t) \exp(-1/\sqrt{\epsilon_{F}t})  \label{eq:invroot}
\end{equation} 
As we are interested in long time behavior, the contribution from the
phase due to the space part of the gauge field will have no effect.

	Now let us investigate contributions coming from 
$S_{int}$ (~\ref{eq:interaction}) only.
The usual rules of diagrammatic perturbation theory apply and we consider only
connected diagrams. The lowest order self-energy due to
the interaction between the fluctuating gauge field and the fermion (Figure 3)
is
\be
\Sigma(i\omega,p)=\int(d\nu dk){\cal D}(\nu,k)G(\omega -\nu,p-k)
v_{\alpha}v_{\beta}(\delta_{\alpha \beta}- k_{\alpha}k_{\beta}/k^{2})
\ee
We choose coordinates in $k$-space ,$k_{\parallel}$ and $k_{\perp}$, which
are parallel and perpendicular to the Fermi vector.  We will
see shortly that $|k_{\parallel}| \ll |k_{\perp}|$ over the relevant ranges
of integration, giving 
$v_{\alpha}v_{\beta}(\delta_{\alpha \beta}- k_{\alpha}k_{\beta}/k^{2})
= v^{2}(1- k_{\parallel}^{2}/(k_{\parallel}^{2}+k_{\perp}^{2})) \approx v^{2}$.
In the Matsubara formalism, 
\begin{eqnarray*}
{\cal D}^{-1}(\nu,k)& =& k_{F}|\nu|/2\pi q + \chi_{0}q^{2} + 
v(q)q^{2}/(4\pi)^{2} \\
v(q)&=&2\pi e^{2}/\epsilon q
\end{eqnarray*}
For the case of Coulomb interactions which we consider here at small wave
vectors and in the clean limit, we have ${\cal D}^{-1}(\nu,k) \approx k_{F}|\nu|/2\pi q +  v(q)q^{2}/(4\pi)^{2}$.
At the Fermi surface,
\begin{eqnarray}
\Sigma(i\omega,p)&=&\int(d\nu dk_{\parallel} dk_{\perp})
(i\omega - i\nu + vk_{\parallel})^{-1} v_{F}^2 (k_{F}|\nu|/2\pi q + 
v(q)q^{2}/(4\pi)^{2})^{-1} \\ 
&\approx & \frac{2iv_{F}\epsilon}{\pi e^{2}}\omega \ln |4\epsilon \omega/
e^{2}k_{F}| \label{eq:selfenergy}
\end{eqnarray}
This self-energy is characteristic of marginal Fermi liquids.

Finally, we investigate  perturbative contributions from cross terms between 
the contour phase and the interaction. To order $g^{2}$ we retain
\begin{eqnarray}
\langle \psi_{e}(r,\tau)  \psi_{e}^{\dag}(0,0) \rangle_{cross} & = &
\int D\psi D\psi^{\ast} Da_{i}e^{-S_{0}}
 \psi(t)\psi^{\dag}(0) \nonumber \\
                                                       &   &
[1+i\int_{C_{0}}d\vec{r\prime\prime}\cdot g\vec{a}(\vec{r\prime\prime},0)]
[1-i\int_{C_{r}}d\vec{r\prime}\cdot g\vec{a}(\vec{r\prime},\tau)] \nonumber \\
                                                       &   &
[1 -ig\int d\tau ' d^{2}r'\psi^{*}(r'\tau ')\vec{a}(r',\tau ') \cdot \nabla 
\psi(r',\tau ')] /Z \nonumber \\
							& =  &
[-\int_{C_{r}}dr''_{\mu}\int d\tau 'd^{2}r'{\cal D}_{\mu \alpha}
(r''-r',\tau - \tau ') \nonumber \\			&    &
+ \int_{C_{0}}dr''_{\nu}\int d\tau 'd^{2}r'{\cal D}_{\nu \alpha}
(r''-r',0 - \tau ')] \nonumber \\
							&    &
\langle \psi(r,\tau)  \psi^{\dag}(0,0)\psi^{\dag}(r',\tau ')
\nabla _{\alpha} \psi(r',\tau ')  \rangle
 \label{eq:cross}
\end{eqnarray}
Fourier transforming
\be
G(\omega,k)_{cross}=
G^{0}(\omega,k)\{1+2\int(d\nu dq){\cal D}(\nu,q)G^{0}(\omega -\nu,k-q)[\frac{k_{x}}{q_{x}}
- \frac{\bf{k}\cdot \bf{q}}{q^{2}}]\} + \cdots \label{eq:fcross}
\ee
Notice that the expression is not rotationally invariant, while the physical
system is.  This is due to the convention of measuring angles
with respect to the x-axis.  The non-invariant part will only contribute a
phase to the electron Green's function and this will not affect any measurable
quantity.  When we go to the next order in the perturbation due to cross
terms between the Chern-Simons phase and the interaction, we find that
the diagrams cross.  However, to logarithmic accuracy, they may be decoupled,
meaning that interference can be neglected.  At higher, orders, calculating the
isotropic part is messy.  Instead, we calculate the anisotropic part and see
that the coefficients look like the Taylor series for an exponential.  We
expect the isotropic part to behave the same way.  
Combining (~\ref{eq:selfenergy}) and (~\ref{eq:fcross})  we obtain
\begin{equation}
G(\omega,k)= G^{CS}(\omega,k)\exp[\frac{|\omega|}{8\pi^{2} \epsilon_{F}}
\ln|\epsilon_{F}/\omega| + \frac{i}{\pi^{2}} \ln|k_{F}/\omega|]
\label{eq:green}
\end{equation}
where 
\be
G^{CS}(\omega,k)=1/(G^{o}(\omega,k)^{-1} -\Sigma(i\omega))
\label{eq:gcs}
\ee
The phase of the exponential in (~\ref{eq:green}) is not rotationally invariant
and was evaluated for the case when $k$ lies along the $x$-axis.  As was
mentioned previously, the phase will not be detectable in any physically
measurable quantity.  The real part of the exponent is rotationally invariant. 

However, as $\omega$ goes to zero, the real part vanishes.  The Chern-Simons
phase makes no contribution. For small $\omega$, we are left with $G(\omega,k)
= G^{CS}(\omega,k)$.  The one particle Green's function of the Chern-Simons
fermion is the same as the one-particle Green's function of the physical
electron.  It is as if the quasiparticle can tunnel coherently. 
We may obtain the tunnelling current from the Kubo formula
\be
I(V)=2e\sum_{k,p}|T_{k,p}|^{2}\int_{-\infty}^{\infty}\frac{d\epsilon}{2\pi}
A_{R}(k,\epsilon)A_{L}(p,\epsilon + eV)(n_{F}(\epsilon) - n_{F}(\epsilon+eV))
\label{eq:kubo}
\ee
where $A_{R}$ and $A_{L}$ are the spectral functions for the right and
left planes.  The tunnelling matrix element $T_{k,p}$ is assumed to conserve
transverse momentum.  From the one particle Green's function (~\ref{eq:gcs}), 
the Kubo formula gives
\begin{eqnarray*}
I(V)=e|T_{k,p}|^{2}a^{2}(m/2)\int_{-\infty}^{\infty}d\xi \int_{-eV}^{0}d\omega
\frac{|\omega||\omega+eV|}
{[\omega-\xi-a\ln|b\omega|\omega]^{2} +[a\pi|\omega|/2]^{2}} \nonumber \\
\frac{1}
{[\omega +eV -\xi-a\ln|b(\omega +eV)|(\omega +eV)]^{2} 
+[a\pi|\omega +eV|/2]^{2}} 
\end{eqnarray*}
where $a=2\epsilon v_{F}/\pi e^{2}$ and $b=\pi a/\epsilon _{F}$.
In the dirty case, $\omega \tau \ll 1$, $\Sigma(\omega) \sim
\frac{-i sgn(\omega)}{\tau}\ln |\epsilon_{F}/\omega|$, a larger 
contribution than in the clean case. Experimental measurements
of the relaxation time give $1/\tau \simeq 0.1 meV$\cite{du}.  To logarithmic
accuracy, (~\ref{eq:kubo}) gives
\begin{equation}
I(V)  \propto   \left\{ \begin{array}{ll}
\tau V / \ln^{2}|\epsilon_{F}/V| 
  &    \hspace{1.5in}  \mbox{\ \ \  clean:} \hspace{0.75in}  \mbox{ $ V \ll 1/\tau $} \\
  1/ \ln^{2}|\epsilon_{F}/V|  
  &    \hspace{1.5in}  \mbox{\ \ \ dirty:} \hspace{0.25in}   \mbox{ $ 1/\tau \ll V \ll \epsilon_{F}$}
		\end{array}
		\right.
\end{equation}
This result is completely different from experiments\cite{eis}.
\section{Semi-Classical Approximation}
	 In this section we will argue that the Chern-Simons phase can be
viewed as a current source in an effective Lagrangian of the gauge fields.  
This source term will induce a vector potential, just as in classical
electrodynamics.  From the induced potential, we can write down the force
that a charged particle would feel as it propagates from the origin at time
zero to the point $r_{0}$ at time $t_{0}$.
We will see at the classical level that the induced potential has no effect
on the tunnelling rate.

	We rewrite (~\ref{eq:funcint}) such that the CS phase is part of the 
action.
\be
{\cal L}= \frac{1}{2} a\Pi a - \vec{j}^{CS}\cdot\vec{a}
\, , $$
\ee
where $\vec{j}^{CS}$ is determined by 
\be
-i\int_{C_{r_{0}}- C_{0}}
d\vec{r\prime}\cdot \vec{a}(\vec{r\prime},t_{0})
   = i\int j^{CS}(\vec{r_{0}},t_{0};\vec{r\prime},t\prime)\cdot\vec{a}(\vec{r\prime},t\prime)
     dx\prime dy\prime dt\prime
\label{eq:CS current}
\end{equation}

The current source $\vec{j}^{CS}$ can be written 
in terms of $\delta$-functions and $\theta$-functions to match Figure 1.
For concreteness,  consider the sample to be a square of side length $2l$. 
\begin{eqnarray*}
\vec{j}^{CS}(\vec{r_{0}},t_{0};\vec{r\prime},t\prime) = -\delta(t'-t_{0})g
\{\hat{y}(\delta(x'-l)[\theta(y') -\theta(y'-l)]\\
-\delta(x'+l)[\theta(y'-y_{0}) -\theta(y'-l)]) 
+\hat{x}(\delta(y'-y_{0})[\theta(x'+l) -\theta(x'-x_{0})]\\
-\delta(y'-l)[\theta(x'+l) -\theta(x'-l)]) \} 
+\delta(t'-0)g\{\hat{y}(\delta(x'-l)[\theta(y') -\theta(y'-l)]\\
-\delta(x'+l)[\theta(y') -\theta(y'-l)]) 
+\hat{x}(\delta(y')[\theta(x'+l) -\theta(x')]\\
-\delta(y'-l)[\theta(x'+l) -\theta(x'-l)]) \}
\end{eqnarray*}
	The current source $\bf{j^{CS}}$  makes the electron  feel a
 force, but with negligible effect on the trajectory.  The induced classical
vector potential is given by
\ba
a_{i}^{cl} & = & D_{ij}j^{CS}_{j} 
          \ea
This leads to a time-dependent electric field which gives a time-dependent
force in Newton's equation of motion.  Suppose we are interested in the
propagator of an electron which starts at the origin at $t=0$ and 
travels along the $x$-axis to the
point $x=x_{o}$ at $t=t_{o}$.  Its classical equation of motion is
\be
\ddot{x}=\sqrt{t_{cl}} \{ \frac{x}{t^{5/2}}+\frac{x-x_{o}}{(t_{o}-t)^{5/2}} \}
\ee
where $t_{cl}= 3\hbar ^{2}(4\pi \epsilon p_{F}/e ^{2})^{3/2}/16\pi mp_{F}$.
This is a boundary value problem which can be solved numerically.  The
trajectory is roughly linear except for the neighborhood of the initial
and final points.  In the semiclassical approximation, the propagator is
the zero field propagator times the phase factor of the line integral of
the gauge field along the classical trajectory.  However, this phase will
not affect the tunnelling rate.  Thus, agreeing with our many-body calculation
above, the Chern-Simons phase has no physical effect.
	
\section{Conclusion}
	
We have studied the Green's function of physical electrons in the half-filled 
Landau level.  Although these Green's
functions differ from those of the quasiparticles,
perturbation theory about the mean field solution gives only small 
(logarithmic) corrections to normal metallic properties.  In particular, 
interactions with the gauge field do not open up a gap. This theoretical
result is at odds with experiments in which tunnelling is exponentially 
suppressed at low voltages, implying a pseudo-gap.  One possibility is
that our theoretical model contains the correct physics, but  
perturbative corrections to mean field solution at $\nu = 1/2$ fail to 
capture the physical density fluctuations
that prevent an electron from propagating.  Another possibility is that
we have left some physics out of our model. We now speculate on other 
possible scenarios that might resolve the contradiction between theory and
experiment.
	
We have assumed that, in the absense of tunnelling, the electrons in one
layer are uncorrelated with the electrons in another layer.  Bonesteel
\cite{bone} considered, in addition to the intralayer Coulomb repulsion,
an interlayer Coulomb repulsion, for the case $\nu = 1/4$.  He found
that interlayer correlations lead to an attractive pairing interaction
between fermions in different layers.  This could lead to a 
superconducting gap.  However, there are two arguments against this mechanism
of gap formation.  One is that in-plane superconductivity is not observed.
Another is that the layers are too far apart for the existence of an 
appreciable inter-planar
Coulomb interaction.  Because of the in-plane Coulomb interaction, the density
is approximately constant and the net field of the electrons and positive
background falls of exponentially.

Another possibility is that tunnelling is strongly affected by disorder.
For instance, we have not considered the dynamics of the electron as it moves through
the barrier.  Suppose that the barrier contains a ``trap'' in which the 
electron spends a significant amount of time, $t_{trap}$.  In other
words, in the single electron picture, the electron tunnels in a two
step process: first it moves out of its layer and into the trap, and
then it moves from the trap to the other layer.  In the many body picture,
this process will be suppressed in a fashion similar to the orthogonality
catastrophe in the x-ray edge problem.  When the electron initially gets
trapped below the surface, the electrons remaining in the layer rearrange
themselves in response to an effective localized charge.  The wave function
of the rearranged state is nearly orthogonal to the wave function 
of the original state.  This means that, for  $\omega_{c}t_{trap} \gg 1$,
the matrix element for the electron to move
into the trap is small, and tunnelling is suppressed.  We emphasize that,
\emph{a priori}, there is no reason to believe that tunnelling involves
mid-barrier trapping as an intermediate step.  It is just
another possible avenue to investigate.

\section{Acknowledgements}
I would like to thank Lev Ioffe for patiently introducing me to both the
experimental and theoretical aspects of the half-filled Landau level.  I
have also benefitted from discussions with Piers Coleman, 
David Langreth, Dan Friedan, S.C. Zhang, J. Jain, Andy Schofield,
Vlad Dobrosavlevic, Robin Chatterjee, Eduardo Miranda, Marcelo Rozenberg, 
Juana Moreno, Boris Narozhny, Revaz Ramazashvilli and William Bigel.

\begin{figure}
\caption{Path over which to integrate the vector gauge field.  The shaded
area represents the physical sample.  The thick black line represents the
directed path.  The path is gauge dependent and in this case coresponds
to the convention that angles are measured with respect to the $x$-axis.}
\end{figure}
\begin{figure}
\caption{Lowest order diagrams coming from the Chern-Simons
phase.}
\end{figure}
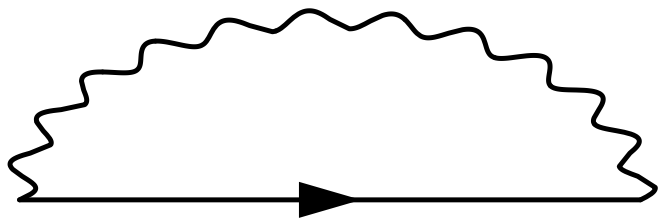
\begin{figure}
\caption{Lowest order self-energy diagram due to the interaction between
the fluctuating gauge field and the fermion.}

\end{figure}


\begin{references}

\bibitem{boebinger} G.S. Boebinger, A.M. Chang, H.L. Stormer and D.C. Tsui
Phys. Rev. Lett. {\bf B 55}, 1606 (1985).
\bibitem{prange} Albert M. Chang, in: The Quantum Hall Effect, Eds. 
Richard E. Prange and Steven M. Girvin (Springer-Verlag, New York, 1987) 
p. 203.  
\bibitem{eis}J.P. Eisenstein, L.N. Pfeiffer, and K.W. West,
Surf. Sci. {\bf 305}, 393 (1994). 
\bibitem{he}Song He, P.M. Platzman and B.I. Halperin,
Phys. Rev. Lett. {\bf 71}, 777 (1993).
\bibitem{efros}A.L. Efros and F.G. Pikus, Phys. Rev. {\bf B 48}, 14694, (1993).
\bibitem{du}R.R. Du, H.L. Stormer, D.C. Tsui, L.N. Pfeiffer, and K.W. West,
Phys. Rev. Lett. {\bf 70}, 2944 (1993).
\bibitem{hlr} B.I. Halperin, P.A. Lee, and N. Read, Phys. Rev. {\bf B 47}, 7312 (1993).
\bibitem{kz}V. Kalmeyer and S.C. Zhang, Phys. Rev. B {\bf 46}, 9889 (1992).
\bibitem{bone}N.E. Bonesteel, Phys. Rev. B {\bf B 48}, 11484, (1993).
\end{references}
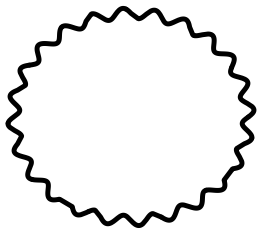
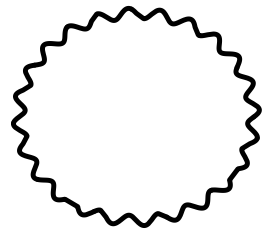
\end{document}